\documentclass[aps,twocolumn,jcp]{revtex4}
\usepackage[utf8x]{inputenc}
\usepackage{graphicx}
\usepackage{latexsym}
\usepackage{amsmath}
\usepackage{amssymb}
\usepackage{bm}

\newcommand{\bee}{\begin{eqnarray}}
\newcommand{\eee}{\end{eqnarray}}
\newcommand{\ba}{\begin{array}}
\newcommand{\ea}{\end{array}}
\newcommand{\bc}{\begin{center}}
\newcommand{\ec}{\end{center}}
\newcommand{\bi}{\begin{itemize}}
\newcommand{\ei}{\end{itemize}}

\begin{document}
\title{Brownian motion of a particle with arbitrary shape}
\author{Bogdan Cichocki}
\email{cichocki@fuw.edu.pl} 
\affiliation{Institute of Theoretical Physics, Faculty of Physics, University of Warsaw, Pasteura 5,
  02-093 Warsaw, Poland}

\author{Maria L. Ekiel-Je\.zewska}
\email{mekiel@ippt.pan.pl}\thanks{Corresponding author}
 \affiliation{Institute of Fundamental Technological Research,
             Polish Academy of Sciences, Pawi\'nskiego 5B, 02-106 Warsaw, Poland}

\author{Eligiusz Wajnryb}
\email{ewajnryb@ippt.pan.pl}
 \affiliation{Institute of Fundamental Technological Research,
             Polish Academy of Sciences, Pawi\'nskiego 5B, 02-106 Warsaw, Poland}

\date{\today}
\begin{abstract}
\bc Abstract \ec
Brownian motion of a particle with an arbitrary shape is investigated theoretically. Analytical expressions for the time-dependent cross-correlations of the Brownian translational and rotational displacements are derived from the Smoluchowski equation. The role of the particle mobility center is determined and discussed.
\end{abstract}

\maketitle
\section{Introduction}
Brownian motion of particles with arbitrary shapes has been recently investigated in many different contexts, including proteins, DNA, nanofibers, actins or other biological nano and micro objects \cite{2D,holografic,leptospira,Adamczyk,Winkler,Goldstein,Kraft}. There is a rapidly growing number of experimental and numerical studies, which give rise to interesting fundamental questions, challenging for theoreticians. 

For nanoparticles, the characteristic time $\tau_R$ of the rotational Brownian diffusion is typically much smaller than the time resolution $t$ in the experiments, $t>>\tau_R$. Therefore, they can be treated as point-like spherical particles, and described by the standard Brownian theory \cite{Kampen}.

 However, for Brownian microparticles, $\tau_R$ is of the order of seconds, and therefore 
non-negligible in comparison to the typical time scales $t$ of the measured Brownian motion, $t \lesssim \tau_R$. 
For example, in Ref. \!\cite{Kraft}, the Brownian motion of a non-symmetric microparticle was investigated experimentally and numerically at the time scales comparable with the characteristic time $\tau_R$ of the rotational Brownian diffusion, 
and in this case the standard approach \cite{Kampen} is not sufficient.  Therefore, the idea was to measure and model numerically the time-dependent cross-correlations of the Brownian translational and orientational displacements of microparticles with different shapes. These results were next also used to determine the mobility and friction matrices.   

The goal of our work is to derive theoretically the explicit analytical expressions for  cross-correlations of the Brownian translational and orientational displacements at all the time scales, and for arbitrarily shaped particles.  

Parts of such a theoretical analysis have been already done. The motivation came from Ref. \cite{Adamczyk}, focused on the problem of determining whether 
measurements of the self-diffusion coefficient and intrinsic viscosity of fibrynogen can be used to determine the protein configuration.  
Evaluation of both these quantities for particles with arbitrary shapes hinges on solving fundamental problems: what is the Brownian contribution to the intrinsic viscosity, and is the self-diffusion coefficient sensitive to the choice of a point on particle for which the mean square displacement is determined.  

The expressions for the intrinsic viscosity of a Brownian non-symmetric particle were theoretically derived in Refs.~\cite{Rallison,CEJW_2012intrinsic}. The relation between the Brownian mean square displacements of different points of a particle was determined in Ref.~\cite{CEJW_2012trans}.

In this work, we consider a single Brownian particle of an arbitrary shape, in general non-isotropic and non-axisymmetric. Starting from the Smoluchowski equation \cite{SE1}-\cite{SE3}, we develop a new formalism, which allows to determine 
the particle rotational and translational motion in a much simpler way as e. g. in Ref. \cite{Favro}, which is based on the Euler angles and Wigner functions. 

Moreover, the essential result of this work is that using our new formalism, we derive simple explicit analytical expressions for the cross-correlations of the Brownian translational and rotational displacements. No such  formulas have been known yet - instead, numerical Brownian simulations have been extensively used, as e. g. in Ref.~\cite{Kraft}.     

\section{System and its theoretical description}
We consider an isolated Brownian particle of arbitrary shape immersed in an unbounded 
fluid of viscosity $\eta$ and temperature $T$. Its state will be described by $X\!=\!(\mathbf{R},%
\widehat{\mathbf{\Omega }})$, where $\mathbf{R}$ is the position of
the chosen particle point and $\widehat{\mathbf{\Omega }}$ describes the
particle orientation, in particular in terms of the Euler angles which
specify the orientation of the particle body-fixed axes with respect to the
space-fixed axes. Probability distribution of finding the particle at time $%
t $ in state $X$ will be denoted by $P(X,t)$. It is normalized~as 
\begin{equation}
\int dX\ P(X,t)=1,
\end{equation}%
with 
\begin{equation}
dX=d\mathbf{R\ }d\widehat{\mathbf{\Omega }},
\end{equation}%
and the element $d\widehat{\mathbf{\Omega }}$ defined in terms of the Euler
angles in the same way as in Ref.~\cite{BernePecora} on page 161.

The evolution of the probability distribution $P(X,t)$ is governed by the
Smoluchowski equation%
\begin{equation}
\frac{\partial }{\partial t}P(X,t)=\mathcal{L}\left( X\right) P(X,t)
\label{SE}
\end{equation}%
with the Smoluchowski operator, which in the absence of external fields has
the form 
\begin{equation}
\mathcal{L}\left( X\right) =\ \mathbf{\nabla }_{X}\text{\/~}\cdot \bm{D}(%
\widehat{\mathbf{\Omega }})\cdot \mathbf{\nabla }_{X}.  \label{SO}
\end{equation}%
In the above equation, 
\begin{equation}
\mathbf{\nabla }_{X}=\left(\frac{\partial }{\partial \mathbf{R}},\frac{\partial }{%
\partial \bm{\alpha }}\right),
\end{equation}%
with 
\begin{eqnarray}
\frac{\partial }{\partial \mathbf{R}}&=&\left( \frac{\partial }{\partial R_1},%
\frac{\partial }{\partial R_2},\frac{\partial }{\partial R_3}\right),
\\
\frac{\partial }{\partial \bm{\alpha }}\text{\/}&=&\left( \frac{%
\partial }{\partial \alpha _{1}},\frac{\partial }{\partial \alpha _{2}},%
\frac{\partial }{\partial \alpha _{3}}\right),
\end{eqnarray}%
where $\alpha _{k}$ is the angle of rotation around 
the axis $k$;
here and in the following, we denote the Cartesian indexes corresponding to $x,y,z$ by $k=1,2,3.$ 
According to the fluctuation-dissipation theorem \cite{Kampen}, the $6\times 6$
diffusion matrix $\bm{D}$ is proportional to the $6\times 6$ mobility
matrix~$\bm{\mu}$,  
\begin{equation}
\bm{ D}(\widehat{\mathbf{\Omega }})=k_{B}T\ \bm{\mu }(\widehat{%
\mathbf{\Omega }}),  \label{Dmu}
\end{equation}

It is important to stress that the matrix $\bm{\mu }$ does not depend on
position $\mathbf{R}$, and its dependence on $\widehat{\mathbf{\Omega }}$
follows entirely from the particle rotation; it means that $\bm{\mu }$
rotates with the particle (see Appendix \ref{A}). 
 As we see from Eq.~\eqref{Dmu}, the
single-particle Brownian motion is determined by its 6$\times $6 mobility
matrix $\bm{\mu }$. If applied to the external force $\mathbf{F}$ and
torque $\mathbf{T}$ exerted on the particle, in the absence of external ambient flows, it results in the particle
translational and rotational velocities, $\mathbf{U}$ and ${\bm \omega}$, 
\begin{equation}
\begin{pmatrix}
\mathbf{U} \\ 
{\bm \omega}%
\end{pmatrix}%
=\bm{\mu }\cdot 
\begin{pmatrix}
\mathbf{F} \\ 
\mathbf{T}%
\end{pmatrix}%
\end{equation}%
The mobility matrix $\bm{\mu }$ consists of four Cartesian 3$\times $%
3 sub-matrices, 
\begin{equation}
\bm{\mu }=%
\begin{pmatrix}
\bm{\mu }^{tt} & \bm{\mu }^{tr} \\ 
\bm{\mu }^{rt} & \bm{\mu }^{rr}%
\end{pmatrix}%
,  \label{Mu}
\end{equation}%
with the indices $t,r$ denoting the translational and rotational components, respectively. The 6$\times $6 mobility matrix $\bm{\mu }$ is symmetric \cite{KimKarrila}, therefore
 $\mu _{ij}^{tt}=\mu _{ji}^{tt}$,\thinspace\ $\mu _{ij}^{rt}=\mu
_{ji}^{tr}$, \thinspace $\mu _{ij}^{rr}=\mu _{ji}^{rr}$. Therefore, the 3$\times $3 matrices $\bm{\mu }^{tt}$ and $\bm{\mu }^{rr}$ are symmetric, but $\bm{\mu }^{tr}$ in general is not symmetric. 
Analogical notation is adopted and the symmetry properties are held for the corresponding 3$\times$3 sub-diffusion matrices, defined with the use of Eq. \eqref{Dmu}.

In general, the mobility matrix $\bm{\mu }$ depends on the choice of a reference center which is observed. 
The hydrodynamic mobility center is such a point for which the
rotational-translational mobility matrix $\bm{\mu }^{rt}$ is symmetric.
The position of this point is explicitly specified e.g. in Refs. \cite{BruneKim:1993} and \cite{KimKarrila}.

Transformation relations between the mobility $\bm{\mu }_{1}$ for the reference 
center $\mathbf{r}_{1}$ and the mobility $\bm{\mu }_{2}$ for the reference center $%
\mathbf{r}_{2}$, called translational theorems for mobility matrices, are the following, 
\begin{eqnarray}
\bm{\mu }_{2}^{rr} &=&\bm{\mu }_{1}^{rr}=\bm{\mu }^{rr},
\label{trans1} \\
\bm{\mu }_{2}^{rt} &=&\bm{\mu }_{1}^{rt}+\bm{\mu }^{rr}\times (%
\mathbf{r}_{2}-\mathbf{r}_{1}),  \label{trans2} \\
\bm{\mu }_{2}^{tt} &=&\bm{\mu }_{1}^{tt}-(\mathbf{r}_{2}-\mathbf{r}%
_{1})\times \bm{\mu }^{rr}\times (\mathbf{r}_{2}-\mathbf{r}_{1})  \notag
\\
&-&(\mathbf{r}_{2}-\mathbf{r}_{1})\times \bm{\mu }_{1}^{rt}+\bm{\mu }%
_{1}^{tr}\times (\mathbf{r}_{2}-\mathbf{r}_{1}),  \label{trans3}
\end{eqnarray}
with the notation that for a second rank tensor $\mathbf{A}$ and a vector $%
\bm{V}$, one has 
\bee
[\mathbf{A}\!\times\! \bm{V}]_{ij}\!\equiv\!
A_{ik}\epsilon _{jkl}V_{l},\hspace{0.4cm} [\bm{V}\!\times \!\mathbf{A}]_{ij}\!\equiv\!
\epsilon _{ilk}V_{l}A_{kj}.\;\;\;\; \label{vecten} 
\eee
 Here and later on, we use the Einstein's summation convention, unless it is explicitly written that the repeating indices are fixed. 

In particular, the translational theorems listed above can be applied to obtain the mobility matrices for an arbitrary reference center, if $\bm{\mu }_c$ for the hydrodynamic mobility center is known. The advantage of using this special center is not only the simplicity of the corresponding expressions. 
In Ref.~\cite{CEJW_2012trans} it has been shown that 
the hydrodynamic mobility center is especially meaningful for the translational diffusion process. In this work, we will show that
such a reference center is also essential for the analysis of the cross-correlations of the Brownian translational or rotational displacements.

Taking into account the structure of the mobility and diffusion matrices, we split the
Smoluchowski operator, defined in Eq.~\eqref{SO}, into four parts,
\begin{equation}
\mathcal{L}=\mathcal{L}^{tt}+\mathcal{L}^{tr}+\mathcal{L}^{rt}+\mathcal{L}%
^{rr},  \label{L1}
\end{equation}
with
\begin{eqnarray}
\mathcal{L}^{tt} &=&\frac{\partial }{\partial R_{k}}D_{kj}^{tt}\frac{%
\partial }{\partial R_{j}},  \label{Ltt}\\
\mathcal{L}^{rr} &=&\frac{\partial }{\partial \alpha _{k}}D_{kj}^{rr}\frac{%
\partial }{\partial \alpha _{j}}\label{Lrr}\\
\mathcal{L}^{rt} &=&\frac{\partial }{\partial \alpha _{k}}D_{kj}^{rt}\frac{%
\partial }{\partial R_{j}},\label{Lrt}\\
\mathcal{L}^{tr}&=&\frac{%
\partial }{\partial R_{k}}D_{kj}^{tr}\frac{\partial }{\partial \alpha _{j}}.\label{Ltr}
\end{eqnarray}
In the above expressions for $\mathcal{L}^{tt}$ and $\mathcal{L}^{rr},$ the order of the diffusion coefficients and both derivatives can be interchanged. This property can be easily derived taking into 
account that $\bm{D}^{tt}$ does 
not depend on ${\bf R}$, and 
$\bm{D}^{rr}$ is symmetric. Below, we will show that for any second rank symmetric tensor~$\mathbf{S}
$,
\begin{equation}
\frac{\partial }{\partial \bm{\alpha }}\cdot \mathbf{S=0}.\label{Azero}
\end{equation}

For the choice of the hydrodynamic mobility center, $\bm{D}^{rt}$ and $\bm{D}^{tr}$ are  also symmetric 
and therefore in Eqs. \eqref{Lrt}-\eqref{Ltr} the order of all the operations is arbitrary.

We will now derive the relation \eqref{Azero}.
Let $\bm{B}$ be a tensor of rank $n$.
Suppose that $\bm{B}$ rotates with the particle, and the rotation matrix $\bm{\mathcal R}$ transforms an initial 
particle orientation $\widehat{\mathbf{\Omega }}_0$ to a final orientation $\widehat{\mathbf{\Omega }}$. 
Then, the components $B_{i_{1}i_{2}...i_{n}}$  of $\bm{B}$, 
 with $i_{k}=1,2,3$ for $%
k=1,...n$,  change 
according to the formula \cite{Rallison},
\begin{equation}
B_{i_{1}i_{2}...i_{n}}(\widehat{\mathbf{\Omega }})=\mathcal{R}_{i_{1}j_{1}}%
\mathcal{R}_{i_{2}j_{2}}...B_{j_{1}j_{2}...j_{n}}(\widehat{\mathbf{\Omega }}_0),
\end{equation}
where $\mathcal{R}_{ij}$ are the components of the 
rotation matrix $\bm{\mathcal R}$.

One can show \cite{Rallison} that
\begin{equation}
\frac{\partial }{\partial \alpha _{k}}B_{i_{1}i_{2}...i_{n}}=\epsilon
_{kj_{1}i_{1}}B_{j_{1}i_{2}...i_{n}}+\epsilon
_{kj_{2}i_{2}}B_{i_{1}j_{2}...i_{n}}+...\label{rotB}
\end{equation}

It follows from the above that for a second rank symmetric tensor~$\mathbf{S}
$ the r.h.s. of Eq. \eqref{rotB} is equal to zero, what means that Eq. \eqref{Azero} is satisfied.

\section{Brownian translational and rotational displacements}
In this section, we will describe Brownian motion of a particle with an arbitrary shape.  
At time $t\!\!=\!\!0$ we choose an arbitrary particle point $\mathbf{R}(0)$ and orientation $\widehat{\mathbf{\Omega }}(0)$, and we trace the positions $\mathbf{R}(t)$ and orientations $\widehat{\mathbf{\Omega }}(t)$ at times~$t$. 

We first analyze 
the time-dependent Brownian 
translational displacements, 
\bee
\Delta \mathbf{R} &=& \mathbf{R}(t) - \mathbf{R}(0).
\eee

We want to evaluate the dynamical cross-correlations,
\bee
\hspace{-0.8cm}\!\!&&\!\!\left\langle \Delta \mathbf{R}(t)\Delta \mathbf{R}%
(t)\right\rangle_0 =\nonumber \\
\hspace{-0.8cm}\!\!&&\!\!\int dX\ (\mathbf{R\!-\!R}_{0})(\mathbf{R\!-\!R}_{0})P(\mathbf{R},%
\widehat{\mathbf{\Omega }}\left\vert \mathbf{R}_{0},\widehat{\mathbf{\Omega }}_{0};t\right).
\label{RRR}
\eee
The average $\langle ... \rangle_0$ 
is taken with respect to the particle positions $\mathbf{R}$ and orientations $\widehat{\mathbf{\Omega }}$ at time $t$, with the conditional probability $P\!\left(\mathbf{R},\widehat{\mathbf{\Omega}}\right.
\left\vert \mathbf{R}_{0},\widehat{\mathbf{\Omega 
}}_{0};t\right) $, which takes into account that at $t=0$ the particle is located at $\mathbf{R}(0)=\mathbf{R}_{0}$, and oriented along 
$\widehat{\mathbf{\Omega }}(0)=\widehat{\mathbf{\Omega }}_{0}$. Actually, the average depends on $\mathbf{R}_{0}$ and $\widehat{\mathbf{\Omega }
}_{0}$ and formally, it should be explicitly denoted as $\langle ... \rangle_{\mathbf{R}_{0},\widehat{\mathbf{\Omega 
}}_{0}}$, but we use a simpler notation $\langle ... \rangle_{0}$. 
The average is taken with respect to the conditional probability, which 
satisfies the Smoluchowski equation \eqref{SE}, i.e. it has the form
\begin{equation}
P\!\left(\mathbf{R},\widehat{\mathbf{\Omega }}\right.\left\vert \mathbf{R}_{0},\widehat{\mathbf{\Omega }}%
_{0};t\right) = 
\exp (\mathcal{L}t)\delta (\mathbf{R-R}_{0})\delta (\widehat{\mathbf{%
\Omega}} -\widehat{\mathbf{\Omega }}_{0}).
\end{equation}

To determine change of the particle orientation $\widehat{\mathbf{\Omega }}$, we 
follow Ref.~\cite{Kraft} and at time $t\!=\!0$ we introduce 
three mutually perpendicular unit vectors $\mathbf{u}^{(p)}(0)=\mathbf{u}^{(p)}_0$ which characterize the particle orientation, 
with $p=1,2,3$, and 
we trace their evolution in time.

The orientation $\mathbf{u}^{(p)}(t)$, $p=1,2,3$, at time $t$ can be interpreted as the result of a rotation matrix $
\bm{\mathcal R}(t)$, acting on the initial orientation $\mathbf{u}^{(p)}(0)$, 
\bee
\mathbf{u}^{(p)}
(t)=
\bm{\mathcal R}(t) \cdot \mathbf{u}^{(p)}(0)
.\label{natR}
\eee
With the use of the initial basis $\mathbf{u}^{(p)}(0)$, the components of the rotation matrix are expressed as ${\cal R}_{pq}=\mathbf{u}^{(p)}(0) \cdot \mathbf{u}^{(q)}(t)$, with $p,q=1,2,3$.

The matrix $\bm{\mathcal R}$ is orthogonal, i.e. $\bm{\mathcal R}^T\!=\!\bm{\mathcal R}^{-1}$, where the superscript $T$ stands for the transposition. We now decompose the rotation matrix into symmetric, $\bm{\mathcal R}^{(s)}$, and antisymmetric (skew-symmetric), $\bm{\mathcal R}^{(a)}$, parts, see Eqs. \eqref{sym}-\eqref{anti},
\bee
\bm{\mathcal R}=\bm{\mathcal R}^{(s)} + \bm{\mathcal R}^{(a)}.
\eee
The antisymmetric part can be used to construct a vector, 
defined as
\bee
\Delta {u}_k(t)=-\frac{1}{2} \epsilon_{kij} {\mathcal R}^{(a)}_{ij}(t).\label{DeltauRa}
\eee
This  vector is 
parallel to the Euler rotation axis \cite{Euler}, with its length  
equal to the absolute value of the sinus of the rotation angle around this axis. From Eq.~\eqref{natR} it follows that
\begin{equation}
\Delta \mathbf{u}(t)=\frac{1}{2}\sum_{p=1}^{3}\mathbf{u}%
^{(p)}(0)
\times \mathbf{u}^{(p)}(t).\label{defDeltau}
\end{equation}%

Following Ref. \cite{Kraft}, in this paper we use $\Delta \mathbf{u}(t)$ to describe the rotational Brownian motion. It is good to keep in mind that the antisymmetric part $\bm{{\mathcal R}}^{(a)}$ is not sufficient to describe all the properties of the rotation \cite{Euler}. 

The conditional probability of the occurrence of a given rotation has the form,
\bee
P\!\left(\widehat{\mathbf{\Omega}}\;
\right\vert \!\left.\widehat{\mathbf{\Omega }}_{0};t\right)=
\int d\mathbf{R}\; P\!\left(\mathbf{R},%
\widehat{\mathbf{\Omega }}\;\right\vert \!\left.\mathbf{R}_{0},\widehat{\mathbf{\Omega }}_{0};t\right).\label{redP}
\eee
It satisfies the 
Smoluchowski equation
with the initial condition 
$ P(\widehat{\mathbf{\Omega }},t=0)=\delta(\widehat{\mathbf{\Omega }}-\widehat{\mathbf{\Omega }}_0).$ Therefore,
\begin{equation}
P\left(\mathbf{\widehat{\Omega }}\right.\left\vert \widehat{\mathbf{\Omega }}_{0};t\right) =\exp (\mathcal{L}%
^{rr}t)\delta (\mathbf{\widehat{\Omega} -\widehat{\Omega }}_{0})\label{POmega}
\end{equation}

The conditional probability \eqref{redP} depends only on the rotation matrix $\bm{\mathcal R}$, which leads from $\widehat{\mathbf{\Omega }}_0$ to $\widehat{\mathbf{\Omega }}$ (it does not depend on specific initial  
orientation $\widehat{\mathbf{\Omega }}_0$).  
The Smoluchowski operator ${\cal L}^{rr}$ is self-adjoint \cite{Kampen}, and the forward and backward Smoluchowski equations are the same. Therefore, 
the probability associated with $\bm{\mathcal R}$ is equal to the probability associated with $\bm{\mathcal R}^{-1}$. Saying it differently, this property follows from the detailed balance condition.  \cite{Kampen}

Taking this property into account, we conclude that the symmetric and antisymmetric parts of the rotation matrix are not correlated,
\bee
\langle \bm{\mathcal R}^{(s)}(t) \bm{\mathcal R}^{(a)}(t)\rangle_0 = 0.\label{asde}
\eee
To derive Eq.~\eqref{asde}, we
decompose $\bm{\mathcal R}^{-1}=\bm{\mathcal R}^{T}$ into symmetric and antisymmetric parts,
\bee
\bm{\mathcal R}^{-1}=\bm{\mathcal R}^{(s)} - \bm{\mathcal R}^{(a)},
\eee
and use the relations \bee \hspace{-0.3cm}&&\langle \bm{\mathcal R}^{(s)}(t) \bm{\mathcal R}^{(a)}(t)\rangle_0 = \langle \left( \bm{\mathcal R}^{-1}\right)^{(s)}\!(t) \left(\bm{\mathcal R}^{-1}\right)^{(a)}\!(t)\rangle_0 \nonumber \\\hspace{-0.3cm} &&= - \langle \bm{\mathcal R}^{(s)}(t) \bm{\mathcal R}^{(a)}(t)\rangle_0 \eee

In Ref. \cite{Kraft}, the Brownian motion was described based on 
measurements of the time-dependent 6 x 6 cross-correlation matrix,  
\bee
{\mathbf C}(t) = \left[\!\!  \ba{c} \left\langle \Delta \mathbf{R}(t)\Delta \mathbf{R}
(t)\right\rangle_0 \; \left\langle \Delta \mathbf{R}%
(t)\Delta \mathbf{u}(t)\right\rangle_0\\ 
\left\langle \Delta \mathbf{u}(t)\Delta \mathbf{R}%
(t)\right\rangle_0 \; \,\left\langle \Delta \mathbf{u}(t)\,\Delta \mathbf{u}%
(t)\right\rangle_0
\ea \!\!\right].\label{C(t)}
\eee 
The diffusion matrix $\bm{D}$ was determined as the time-derivative of the correlation matrix ${\mathbf C}(t)$ at time $t=0$,
\bee
\frac{1}{2} \left[ \frac{d}{dt}{\mathbf C}(t)\right]_{t=0} = \bm{D},\label{CD}
\eee
with the components, in analogy to Eq.~\eqref{Mu}, denoted as,
\bee
\bm{D}=\left[\!\!\ba{c}\bm{D}^{tt}\;\bm{D}^{tr}\\\bm{D}^{rt}\; \bm{D}^{rr}
\ea \!\!  \right].
\eee 

The relation Eq. \eqref{CD}  follows from the Smoluchowski equation \eqref{SE}.

In this work, we will describe Brownian motion of a particle with an arbitrary shape by 
deriving explicit analytical expressions for the cross-correlation matrix ${\mathbf C}(t)$ at all times $t$. In general, ${\mathbf C}(t)$ depends on the choice of the reference center. The transformation relations for Eq.~\eqref{C(t)} to another reference center depend on both the antisymmetric and symmetric parts of the rotation matrix, $\bm{\mathcal R}^{(a)}$ and $\bm{\mathcal R}^{(s)}$, respectively. 
These relations will be discussed in Appendix \ref{shift}.

\section{Evolution due to rotational diffusion}
\subsection{Tensors which rotate with the particle}
While evaluating time-dependent cross-correlations, it is important to know how to determine changes due to the rotational diffusion. 
In this case, 
it is sufficient to average with respect to $P(\widehat{\mathbf{\Omega}}
\vert \widehat{\mathbf{\Omega }}_{0};t)$, given by Eq.~\eqref{POmega}. 

Therefore, we consider a  tensor $\mathbf{B}$ which rotates with the particle and assume that $\widehat{%
\mathbf{\Omega }}(t\!=\!0)\!=\!\widehat{\mathbf{\Omega }}_{0}$. For the conditional average of this tensor we find 
\begin{eqnarray}
\left\langle \mathbf{B}(t)\right\rangle_0  &=&\int d\widehat{\mathbf{\Omega }}%
\ \mathbf{B}(\widehat{\mathbf{\Omega }})~P(\widehat{\mathbf{\Omega }}
\left\vert \widehat{\mathbf{\Omega }}_{0};t\right)  \nonumber  \\
&=&\int d\widehat{\mathbf{\Omega }}\ \mathbf{B}(\widehat{\mathbf{\Omega }}%
)\exp (\mathcal{L}^{rr}t)\delta (\widehat{\mathbf{\Omega }}\mathbf{-}%
\widehat{\mathbf{\Omega }}_{0})  \nonumber  \\
&=&\int d\widehat{\mathbf{\Omega }}\ \left\{ \exp (\mathcal{L}^{rr}t)%
\mathbf{B}(\widehat{\mathbf{\Omega }})\right\} \delta (\widehat{\mathbf{%
\Omega }}\mathbf{-}\widehat{\mathbf{\Omega }}_{0})~  \nonumber \\      
&=&\exp (\mathcal{L}^{rr}t)\mathbf{B}(\widehat{\mathbf{\Omega }})\left\vert
_{\widehat{\mathbf{\Omega }}=\widehat{\mathbf{\Omega }}_{0}}\right. \label{BBB}
\end{eqnarray}

The explicit form of the above average depends on the tensor rank. It can be relatively easy evaluated in 
the body-fixed frame in which the rotational diffusion tensor $\bm{D}^{rr}$ is diagonal,
\bee
D^{rr}_{ij}=D_i \delta_{ij}, \hspace{0.5cm} \mbox{for given }i,j=1,2,3.\;\;\label{dia}
\eee
From now on, this frame will be used until the end of the paper. 

The time-dependence of a tensor $\mathbf{B}$ follows from solving the eigenproblem for the Smoluchowski operator ${\cal L}^{rr}$, defined in Eq. \eqref{Lrr}. The eigenvalues depend on the rank of this tensor. 
For example, when the tensor rank is one (i.e. when $\mathbf{B}$ is a vector), we find 
three eigenvalues $-f_{i }^{(1)}$ (as in Ref.~\cite{CEJW_2012trans}), with
\bee
f_{i }^{(1)} &=&3D-D_{i}, \hspace{0.5cm}i =1,2,3,
\label{f1}
\eee
where
\begin{equation}
D=\frac{1}{3}(D_{1}+D_{2}+D_{3}).\label{trrr}
\end{equation}%
Keep in mind that in this work, $D_i$ and $D$ always refer to the rotational-rotational diffusion (we skip the superscript $^{rr}$ for simplicity). 

The corresponding eigenvectors are parallel to the principal axes of the operator $\bm{D}^{rr}$. Denoting by ${V}_{i}(0)$ the components of this vector 
in the frame of reference in which $\bm{D}^{rr}$ is diagonal at $t=0$, and taking into account Eq.~\eqref{BBB}, we obtain, 
\bee
\langle{V}_{i}(t)\rangle_0 =  \exp (-f_{i}^{(1)}t) {V}_{i}(0).\label{Vt}
\eee

For a second rank tensor $\bm{H}(t)$, the problem was solved in Ref.~\cite{Rallison}, and the results are listed in Appendix~\ref{srt}. The
antisymmetric part of $\bm{H}(t)$  may be associated with a vector, and therefore its evolution follows from Eq. \eqref{Vt}, 
with the same eigenvalue $-f_i^{(1)}$, given by Eq.~\eqref{f1}. The evolution of the traceless symmetric part of $\bm{H}(t)$ is associated with five other   eigenvalues, $-f_{i }^{(2)}$, $-f^{(+)}$ and $-f^{(-)}$ as the characteristic exponents in the time decay, where
\begin{eqnarray}
f_{i }^{(2)} &=&3(D_{i }+D),\text{ \ \ \ \ \ }i =1,2,3, \label{f2}\\
f^{(+)} &=&6D+2\Delta , \label{f+}\\
f^{(-)} &=&6D-2\Delta ,\label{f-}
\end{eqnarray}
with $D_{i}$ and $D$ defined in Eqs. \eqref{dia} and \eqref{trrr} and
\begin{equation}
\Delta =\sqrt{%
D_{1}^{2}+D_{2}^{2}+D_{3}^{2}-D_{1}D_{2}-D_{1}D_{3}-D_{2}D_{3}}.
\end{equation}
The trace of $\bm{H}(t)$ is a scalar and as such, it does not depend on time.

The expressions for the time-dependence of the average $\langle \bm{H}(t)\rangle_0$, analogical to Eq.~\eqref{Vt}, are listed in Appendix~\ref{srt} in Eqs.~\eqref{deH}-\eqref{Hsa}.
The important outcome of our analysis is that these expressions 
contain {\it different} characteristic exponents, listed in Eqs.~\eqref{f1} and \eqref{f2}-\eqref{f-}. 

\subsection{Time scales of the rotational diffusion}
From our analysis, outlined in the previous subsection, it follows that in general, there is no single characteristic time scale $\tau_R$ of the rotational self-diffusion.
The exponents, listed in Eqs.~\eqref{f1} and \eqref{f2}--\eqref{f-}, 
determine several characteristic time scales of the translational and rotational correlations,
\bee
&\!&\tau^{(+)}=1/f^{(+)},\;\;\;\;\tau^{(-)}=1/f^{(-)},\nonumber \\
&\!&\tau_i^{(1)}=1/f_i^{(1)},\;\;\;\;\tau_i^{(2)}=1/f_i^{(2)},
\eee
 with $i=1,2,3$. In general, these scales differ from each other, and a careful analysis of the time-dependence at long times in needed.

It is always true that  $\tau_i^{(1)}\!\ge \!\tau_i^{(2)}$ and 
$\tau^{(-)}\!\ge \!\tau^{(+)}$, but $\tau_i^{(1)}$ can be smaller or larger than $\tau^{(-)}$, depending on the particle geometry. 

\section{Mean particle displacement}

In general, due to the rotational diffusion, the mean particle displacement 
is not equal to zero i.e. if $\mathbf{R}$ is the position of the chosen
particle point $\left\langle \bigtriangleup \mathbf{R}(t)\right\rangle_0
=\left\langle (\mathbf{R}(t)-\mathbf{R}(0))\right\rangle_0 \neq 0$. To
find this mean displacement let's first consider special case when the
particle point is the mobility center $\mathbf{R}_{C}$. 

Let's calculate the time derivative of $\langle \bigtriangleup \mathbf{R}_{C}(t)\rangle_0$. Taking into account the Smoluchowski equation \eqref{SE} we have
\bee
\hspace{-0.7cm} &&\frac{d}{dt}\left\langle \bigtriangleup \mathbf{R}_{C}(t)\right\rangle_0 =\nonumber\\
\hspace{-0.7cm} &&\int
dX\ (\mathbf{R}_{C}\mathbf{-R}_{C0})\mathcal{L}P\left(\mathbf{R}_{C},\widehat{\mathbf{%
\Omega }}\right.\left\vert \mathbf{R}_{C0},\widehat{\mathbf{\Omega }}_{0};t\right).\;\;
\eee

Now we will use the property, that for the mobility reference center, all the derivative can be shifted left.
Then, integrations by parts 
give that the above
derivative is equal to zero. Thus, 
\begin{equation}
\left\langle \bigtriangleup \mathbf{R}_{C}(t)\right\rangle_0 =0.\label{DeltaRc}
\end{equation}

Now consider the mean displacement for arbitrary chosen particle point $%
\mathbf{R}$. The vector $\mathbf{R}(t)-\mathbf{R}_{C}(t)$ rotates with the particle. Thus from Eq. \eqref{Vt}
we obtain
\begin{equation}
\left\langle (\mathbf{R}(t)-\mathbf{R}_{C}(t))\right\rangle_{0,i }=(%
\mathbf{R}(0)-\mathbf{R}_{C}(0))_{i }\ \exp (\mathcal{-}f_{i
}^{(1)}t).
\end{equation}
and with (\ref{DeltaRc}) we get
\begin{equation}
\left\langle \Delta \mathbf{R}(t)\right\rangle_{0,i }=(\mathbf{R}_{C}(0)-%
\mathbf{R}(0))_{i }\left\{ 1-\exp (\mathcal{-}f_{i
}^{(1)}t)\right\} \label{evR}
\end{equation}

From Eq. \eqref{evR} it follows that for long times, the mean position of an arbitrary point of the particle tends to the same limit: the initial position of the mobility center,
\bee
\left\langle \mathbf{R}(t)\right\rangle_0 \rightarrow \mathbf{R}_C(0), \hspace{0.5cm}\mbox{for }t\rightarrow \infty.
\eee
This result can be used to experimentally determine the particle mobility center, what is especially useful if the the location of $R_C(0)$ is not known in an analytical form. 

In sections \ref{cpd}-\ref{rtc}, we will follow the motion of the mobility center, $\mathbf{R}(t)\equiv\mathbf{R}_C(t)$, chosen as the reference center in the mobility matrix, see  Eqs.~\eqref{trans1}\nobreak-\nobreak\eqref{trans3}.

\section{Translational-translational correlations}\label{cpd}
\subsection{General expressions}
Let us now consider correlations of a Brownian particle displacements, 
$\left\langle \Delta \mathbf{R}_C(t)\Delta \mathbf{R}_C%
(t)\right\rangle_0$. We remind that we stay in the reference frame in which $\bm{D}^{rr}$ is diagonal and follow the motion of the mobility center $\mathbf{R}_C$.

It is convenient to calculate first the time derivative of the expression \eqref{RRR}:
\bee
\hspace{-0cm} \frac{d}{dt}\left\langle \Delta \mathbf{R}_C(t)\Delta \mathbf{R%
}_C(t)\right\rangle_0 =\nonumber\\
\hspace{-0cm} \int\!\!\! dX (\mathbf{R}_{C}\!\!-\!\!\mathbf{R}_{C0})(\mathbf{R}_C\!\!-\!\!\mathbf{R}_{C0})\,\mathcal{L}\,P\!\left(\!%
\mathbf{R}_C,\widehat{\mathbf{\Omega }}\right. \!\left\vert \mathbf{R}_{C0},\widehat{\mathbf{\Omega }}%
_{0};t\!\right)\!\!\!\!\!\!\!\nonumber\\
\eee

Taking into account the explicit expressions \eqref{Ltt}-\eqref{Ltr} for the Smoluchowski operator and interchanging the order of derivatives we obtain,
\begin{eqnarray}
\hspace{-0.7cm}&&\frac{d}{dt}\left\langle \Delta \mathbf{R}_C(t)\Delta \mathbf{R%
}_C(t)\right\rangle_0 = 2 \left\langle \bm{D}^{tt}(t)\right\rangle_0.
\end{eqnarray}

We perform the above average using Eq. \eqref{BBB}. Since $\bm{D}^{tt}$ is a symmetric tensor we get from Eq. \eqref{Htd} the
following result for diagonal components,
\bee
\hspace{-0.7cm}&&\frac{1}{2}\frac{d}{dt}\left\langle \Delta \mathbf{R}_C%
(t)\Delta \mathbf{R}_C(t)\right\rangle _{0,ii }=\nonumber \\
\hspace{-0.7cm}&&D^{t}+\text{e}%
^{-f^{(-)}t}D_{ii}^{tt\text{ }(-)}+\text{e}^{-f^{(+)}t}D_{ii }^{tt\text{ }(+)}\label{49}
\eee
where $i=1,2,3$ and
\begin{equation}
D^{t}=\frac{1}{3}\text{tr}\bm{D}^{tt}. 
\end{equation}
The meaning of the other symbols is explained in Appendix \ref{srt}. 
After integration of Eq.~\eqref{49} with respect to $t$ one obtains for $i=1,2,3$, 
\bee
\hspace{-0.7cm}&&\frac{1}{2}\left\langle \Delta \mathbf{R}_C(t)\Delta \mathbf{R}_C%
(t)\right\rangle _{0,ii }= D^{t} t+\frac{1\!-\!\text{e}^{-f^{(-)}t}}{%
f^{(-)}}D_{ii }^{tt\text{ }(-)}  \nonumber \\
\hspace{-0.7cm}&&+\frac{1\!-\!\text{e}^{-f^{(+)}t}}{%
f^{(+)}}D_{ii }^{tt\text{ }(+)}.\label{RRii}
\eee

The off-diagonal components $i\ne j$ are given by
\bee
&&\frac{1}{2}\frac{d}{dt}\left\langle \Delta \mathbf{R}_C%
(t)\Delta \mathbf{R}_C(t)\right\rangle _{0,ij }=
\text{e}%
^{-f_{k }^{(2)}t}D_{ij }^{tt},\text{ \ \ \ }, \label{ijne} 
\eee
where $k$ is the remaining third index, $k\ne i,j$. After integration of Eq.~\eqref{ijne} with respect to $t$ one obtains
\bee
&\hspace{-0.5cm}&\frac{1}{2}\left\langle \Delta \mathbf{R}_C(t)\Delta \mathbf{R}_C%
(t)\right\rangle _{0,ij }=\frac{1\!-\!\text{e}^{-f_{k }^{(2)}t}}{%
f_{k }^{(2)}}D_{ij }^{tt}
. \label{RRij}
\eee

In the limit of $t \rightarrow 0$, 
the time derivative of the expressions \eqref{RRii}-\eqref{RRij} approaches the corresponding elements of the diffusion matrix, 
 in agreement with  Eq. \eqref{CD}.

From the relation \eqref{RRii}, taking into account that ${\bm D}^{tt\text{ }(+)}$ and ${\bm D}^{tt\text{ }(-)}$ are traceless, one derives 
the expression for the mean square displacement,
\begin{equation}
\frac{1}{6}\left\langle \Delta \mathbf{R}_C(t)\cdot \Delta 
\mathbf{R}_C(t)\right\rangle= D^{t}t.\label{msd}
\end{equation}
In Eq.~\eqref{msd}, the subscript ``0'' associated with the averaging has been omitted, because the mean square displacement is a scalar, and therefore it does not depend on the orientation, and in particular, on the initial orientation $\widehat{\mathbf{\Omega}}_0$. 

The mean square displacement of the mobility center, given in Eq. \eqref{msd}, reproduces the result from Ref. \cite{CEJW_2012trans}.

For long times, Eqs.~\eqref{RRii} and \eqref{RRij} take the form,
\bee
\hspace{-0.7cm}&&\frac{1}{2}\left\langle \Delta \mathbf{R}_C(t)\Delta \mathbf{R}_C%
(t)\right\rangle _{0,ii }= D^{t} t+\frac{1}{%
f^{(-)}}D_{ii }^{tt\text{ }(-)}  \nonumber \\
\hspace{-0.7cm}&&+\frac{1}{%
f^{(+)}}D_{ii }^{tt\text{ }(+)} + {\cal O}(e^{-t/\tau^{(-)}}),\\
\hspace{-0.7cm}&&\frac{1}{2}\left\langle \Delta \mathbf{R}_C(t)\Delta \mathbf{R}_C%
(t)\right\rangle _{0,ij }=\frac{1}{%
f_{k }^{(2)}}D_{ij }^{tt} + {\cal O}(e^{-t/\tau_k^{(2)}}).\nonumber \\
\eee
It is important to emphasize that there are constant non-vanishing terms in the above expressions. 

\subsection{Special cases}
For an axisymmetric particle, a frame is chosen where the rotational-rotational diffusion matrix is diagonal, with the coefficients
\begin{eqnarray}
D_{1} =D_{2}\neq D_{3}. 
\end{eqnarray}
The translational-translational diffusion matrix is also diagonal,
\bee
D_{ij }^{tt} \!&\!=\!&\!D_{i}^{t}\delta _{ij},\text{ for }i,j=1,2,3,\\
D_{1}^{t}\!&\!=\!&\!D_{2}^{t}\neq D_{3}^{t}.
\eee
The diagonal correlations have the form,
\bee
\hspace{-0.7cm}&&\frac{1}{2}\left\langle \Delta \mathbf{R}_C(t)\Delta \mathbf{R}_C%
(t)\right\rangle _{0,ii}=\nonumber \\
\hspace{-0.7cm}&&D^{t}~t+\frac{1-\text{e}^{-6D_{1}t}}{%
18 D_{1}}(D_{1}^{t}-D_{3}^{t}),
\eee
for $i$=1,2, and 
\bee
\hspace{-0.7cm}&&\frac{1}{2}\left\langle \Delta \mathbf{R}_C(t)\Delta \mathbf{R}_C%
(t)\right\rangle _{0,33}=\nonumber \\
\hspace{-0.7cm}&&D^{t}~t-\frac{1-\text{e}^{-6D_{1}t}}{%
9 D_{1}}(D_{1}^{t}-D_{3}^{t}).
\eee
The off-diagonal components vanish.

For a spherical particle, the rotational and translational diffusion tensors are isotropic,
\bee
D_{1}=D_{2}=D_{3}=D,\\
D_{1}^t=D_{2}^t=D_{3}^t=D^t,
\eee and
\begin{equation}
\frac{1}{2}\left\langle \Delta \mathbf{R}_C(t)\Delta \mathbf{R}_C%
(t)\right\rangle_0 =D^{t}t\text{ }\mathbf{I},
\end{equation}
where $\mathbf{I}$ is the identity tensor.

\section{Rotational-rotational correlations}\label{rrc}
\subsection{General expressions}

From Eq. \!\!\eqref{defDeltau} it follows that 
the rotational-rotational correlations 
read, %
\bee
\hspace{-0.8cm}&&\left\langle \Delta \mathbf{u}(t)\Delta \mathbf{u}%
(t)\right\rangle_0 =\nonumber \\
\hspace{-0.8cm}&&-\frac{1}{4}\!\sum_{p=1}^{3}\sum_{q=1}^{3}%
\mathbf{u}^{(p)}_0\times \left\langle \mathbf{u}^{(p)}(t)\mathbf{u}^{(q)}(t)\right\rangle_0 \times \mathbf{u}^{(q)}_0.
\eee
The average $\langle...\rangle_0$ in the above equation is evaluated based on the expressions provided in Appendix~\ref{srt}.

The off-diagonal elements vanish, and
the diagonal elements, 
$i \!=\!1,2,3,$ are given by
\begin{eqnarray}
\hspace{-0.7cm}&&\left\langle \Delta\mathbf{u}(t)\Delta\mathbf{u}(t)\right\rangle _{0,ii } \!=
\frac{1}{6}-\frac{%
3(D\!-\!D_{i })\!+\!\Delta }{12\Delta }\text{e}^{-f^{(-)} t}
\nonumber \\
\hspace{-0.7cm}&&
-\frac{3(D_{i }\!-\!D)\!+\!\Delta }{%
12\Delta }\text{e}^{-f^{(+)}t}\!
-\!\frac{1}{4}\text{e}^{-f_i^{(2)}
t}
\!+\!\frac{1}{4}\text{e}^{-f_i^{(1)}
t}.\;\;\label{rrii}
\end{eqnarray}

In the limit of $t \rightarrow 0$, 
the time derivative of the above expressions approaches the corresponding elements of the diffusion matrix, 
 in agreement with  Eq. \eqref{CD}.

For long times, 
\bee
\lim_{t\rightarrow \infty} \left\langle \Delta \mathbf{u}(t )\Delta \mathbf{u}%
(t )\right\rangle _{0}&=&\frac{1}{6}\mathbf{I}.
\eee
This result can be also obtained as the average of Eq.~\eqref{rrii} with the equilibrium distribution.

\subsection{Special cases}
For an axisymmetric particle, $D_{1}\!\!=\!\!D_{2}\!\!\neq \!\!D_{3}$, and
\bee
\hspace{-0.7cm}&&\left\langle \Delta\mathbf{u}(t)\Delta\mathbf{u}(t)\right\rangle _{0,11}\!=\!\left\langle \Delta \mathbf{u}%
(t)\Delta \mathbf{u}(t)\right\rangle_{0,22}\!=\!\nonumber \\
\hspace{-0.7cm}&&\frac{1}{6}-\frac{1}{6}%
\text{e}^{-6D_{1}t}-\frac{1}{4}\text{e}^{-(5D_{1}\!+\!D_{3})t}+\frac{1}{4}\text{e}^{
-(D_{1}\!+\!D_{3})t},\\
\hspace{-0.7cm}&&\left\langle \Delta\mathbf{u}(t)\Delta\mathbf{u}(t)\right\rangle _{0,33}
=\frac{1}{6}+\frac{1}{12}\text{e}^{-6D_{1}t} -\frac{1}{2}\text{e}^{
-(2D_{1}+4D_{3})t}\nonumber \\
\hspace{-0.7cm}&&+\frac{1}{4}\text{e}^{-2D_{1}t}.
\eee

To describe the rotational self-diffusion of an axisymmetric particle, it is common \cite{SE1} to trace the change of the particle orientation vector along the symmetry axis, in this paper denoted as $\mathbf{u}^{(3)}(t)$. This vector rotates with the particle, and therefore its evolution follows from Eq.~\eqref{Vt}. Using this relation, we reproduce the standard formula \cite{SE1},
\bee
\left\langle \left[ \mathbf{u}^{(3)}(t)- \mathbf{u}^{(3)}(0)\right]^2\right\rangle = 2(1-e^{-2D_1 t}).
\eee

For a spherical particle, $D_{1}\!=\!D_{2}\!=\!D_{3}\!=\!D$, and 
\begin{equation}
\left\langle \Delta \mathbf{u}(t)\Delta 
\mathbf{u}(t)\right\rangle _{0}=\left[\frac{1}{6}-\frac{5}{12}\text{e}^{
-6Dt}+\frac{1}{4}\text{e}^{-2Dt}\right] \mathbf{I}.
\end{equation}
\vspace{0.3cm}

\section{Rotational-translational correlations}\label{rtc}
\subsection{General expressions}
Using  Eq.~\eqref{defDeltau} we have,
\bee
&\!&\!\hspace{-.4cm}\left\langle \Delta \mathbf{u}(t) \Delta \mathbf{R}_C (t) %
\right\rangle_0 \!=\!
\frac{1}{2}\sum_{p=1}^3 \!\mathbf{u}^{(p)}_0 \!\!\times \!\left\langle \! \mathbf{u}^{(p)}(t) \Delta \mathbf{R}_C (t) %
\!\right\rangle_{\!0}\!.\!\!\!\nonumber \\ 
\eee%
where, as before, $\mathbf{u}^{(p)}_0$ with $p=1,2,3$ are the unit vectors corresponding to the orientation $\widehat{\mathbf{\Omega}}_{0}$, and we choose the body-fixed frame in which the rotational diffusion tensor $\bm{D}^{rr}$ is diagonal. %

As shown in Appendix \ref{A},
\bee
\hspace{-0.3cm}\left\langle  \!\mathbf{u}^{(p)}(t) \Delta \mathbf{R}_C (t) %
\!\right\rangle_0 \!=\!-2 \!\int_0^t\!\! d\tau \,e^{-f_p^{(1)}(t-\tau)}\left\langle\! {\bf A}^{(p)}(\tau)\!\right\rangle_{\!0}\!,\nonumber \hspace{-0.8cm}\\\label{78}
\eee
where the second rank tensor ${\bf A}^{(p)}$ is given by
\bee
{\bf A}^{(p)} &=& \mathbf{u}^{(p)} \times \bm{D}^{rt}.
\eee

We now decompose the tensor ${\bf A}^{(p)}$ and determine  
$\langle {\bf A}^{(p)}(t)\rangle_{0}$
for all its parts as described in Appendix \ref{srt}. The resulting expressions are plugged into Eq.~\eqref{78}, and the integration with respect to $\tau$ is performed.

We obtain the following diagonal Cartesian components of the rotational-translational correlations, 
\begin{eqnarray}
\hspace{-0.5cm}&&\left\langle  \Delta \mathbf{u}(t) \Delta \mathbf{R}_C (t)%
\right\rangle _{0,ii} 
=-\frac{D_{ii }^{rt}\!\!-\!\!D_{jj }^{rt}}{8}~\frac{\text{e}%
^{-f_{k }^{(2)}t}\!-\!\text{e}^{-f_{k }^{(1)}t}}{D_{k }}\nonumber \\
\hspace{-0.5cm}&&- \frac{%
D_{ii }^{rt}\!\!-\!\!D_{kk }^{rt}}{8}~\frac{\text{e}%
^{-f_{j }^{(2)}t}\!-\!\text{e}^{-f_{j}^{(1)}t}}{D_{j }} 
+\frac{D_{ii }^{rt}\!\!+\!\!D_{jj }^{rt}}{2}~t\;\text{e}%
^{-f_{k }^{(1)}t}\nonumber \\
\hspace{-0.5cm}&&+\frac{D_{ii }^{rt}\!\!+\!\!D_{kk
}^{rt}}{2}~t\;\text{e}^{-f_{j }^{(1)}t},\label{rtaa}
\end{eqnarray}%
where $i=1,2,3$ and $j, k $ are 
the remaining second and third indices such that $j +k
=6-i $.

The off-diagonal Cartesian components are, 
\begin{widetext}
\begin{eqnarray}
&&\left\langle  \Delta\mathbf{u}(t) \Delta\mathbf{R}_C(t)\right\rangle _{0,ij } 
=-\frac{D_{ij }^{rt}}{2}\left[ \frac{D_{j }\!-\!D_{k
}\!+\!\Delta }{\Delta }~\frac{\text{e}^{-f^{(-)}t}\!-\!\text{e}^{-f_{k }^{(1)}t}%
}{f^{(-)}\!-\!f_{k }^{(1)}}~+~\frac{D_{k }\!-\!D_{j }\!+\!\Delta }{%
\Delta }~\frac{\text{e}^{-f^{(+)}t}\!-\!\text{e}^{-f_{k }^{(1)}t}}{%
f^{(+)}\!-\!f_{k}^{(1)}}\right. \nonumber\\
&&\left. +~\frac{\text{e}^{-f_{i }^{(2)}t}\!-\!\text{e}^{-f_{j }^{(1)}t}%
}{f_{i }^{(2)}\!-\!f_{j }^{(1)}}~+~\frac{\text{e}^{-f_{i }^{(1)}t}-%
\text{e}^{-f_{j }^{(1)}t}}{f_{i }^{(1)}\!-\!f_{j }^{(1)}}\right] ,\hspace{7cm}i \ne j,\label{rtab}
\end{eqnarray}%
\end{widetext}
where the remaining third index $k =6-i-j $. The expression \eqref{rtab} is not symmetric in $ij $, but its derivative at $t=0$ is symmetric.

In the limit of $t \rightarrow 0$, 
the time derivative of the above expressions approaches the corresponding elements of the diffusion matrix, 
 in agreement with  Eq. \eqref{CD}.
For long times,
\begin{equation}
\lim_{t\rightarrow \infty} \left\langle  \Delta \mathbf{u}(t ) \Delta \mathbf{R}_C%
(t )\right\rangle =0. 
\end{equation}

\subsection{Special cases}
For an axisymmetric particle, $\bm{D}^{tr}$ is skew-symmetric \cite{KimKarrila}. Therefore, it vanishes if evaluated with respect to the mobility center. Taking this into account, we obtain from Eqs. \eqref{rtaa}-\eqref{rtab} the simple result,
\begin{eqnarray}
&&\left\langle  \Delta\mathbf{u}(t) \Delta\mathbf{R}_C(t)\right\rangle _{0} =0.
\end{eqnarray}

\section{Conclusions}
In this work, we performed theoretical analysis of the Brownian motion of a particle with an arbitrary shape. 
We derived analytical expressions for the time-dependent cross-correlations of the displacements of the particle position and orientation. These results were written in the frame of reference in which the rotational-rotational diffusion tensor is diagonal at $t=0$, and with the choice of the mobility center as the reference center. These results can be compared with experimental data, using the following procedure. 

Based on measurements of the time-dependent orientation $\mathbf{u}^{(p)}(t)$, it is possible to determine the correlation tensor $\langle \Delta\mathbf{u}(t) \Delta\mathbf{u}(t)\rangle_0$ and find its principal axes. From our analysis it follows that  
they coincide with the principal axes of the rotational-rotational diffusion tensor. Therefore, to be compared with our theoretical expressions, the experimental data should be recalculated to this new frame of reference in which $\langle \Delta\mathbf{u}(t) \Delta\mathbf{u}(t)\rangle_0$ is diagonal. 

The initial position of the mobility center $\mathbf{R}_C(0)$ can be determined experimentally by tracing in time the average position $\langle \mathbf{R}(t)\rangle_0$ of any particle point, and using our finding that $\langle \mathbf{R}(t)\rangle_0 \rightarrow \mathbf{R}_C(0)$ when $t \rightarrow \infty$. 

The transformation formulae for $\mathbf{C}(t)$ from the mobility center to another arbitrary point can be derived analytically, as described in Appendix \ref{shift}. However, the transformation formula for $\langle \Delta\mathbf{R}(t) \Delta\mathbf{R}(t) \rangle _{0}$ involves the symmetric part of the rotation matrix $\bm{\mathcal R}^{(s)}$, which cannot be expressed in terms of $\mathbf{C}(t)$.


Our simple explicit analytical expressions for the correlations ${\bf C}(t)$, valid for the mobility reference center, can be compared with the experimental data, measured for an arbitrary reference center, using the following procedure. 
Once the principal axes of the rotational-rotational diffusion tensor are determined, and used as the new coordinate system, with the recalculated time-dependent orientation $\mathbf{u}^{(p)}(t)$, and the initial position of the mobility center $\mathbf{R}_C(0)$ is determined experimentally, 
this information can be used 
as follows. First, the difference between the initial position $\mathbf{R}(0)$ of an arbitrary reference center, traced in experiments, and $\mathbf{R}_C(0)$, 
can be expressed as a linear combination of the orientation vectors $\mathbf{u}^{(p)}(0)$,
\bee
\mathbf{R}(0)-\mathbf{R}_C(0)=\sum_{p=1}^3 a_p \mathbf{u}^{(p)}(0),
\eee
and the coefficients $a_p$ can be determined. Taking into account that $\mathbf{R}(t)-\mathbf{R}_C(t)$ rotates with the particle, we can use the same coefficients $a_p$ to evaluate $\mathbf{R}_C(t)$ as
\bee
\mathbf{R}_C(t)=\mathbf{R}(t)-\sum_{p=1}^3 a_p \mathbf{u}^{(p)}(t),\label{act}
\eee

Eq.~\eqref{act} allows to express  (the unknown) stochastic trajectory $\mathbf{R}_C(t)$ in terms of (the known)  stochastic trajectory $\mathbf{R}(t)$ and orientation $\mathbf{u}^{(p)}(t)$. 
This allows 
to extract from the measured data the correlations of the Brownian displacements of the mobility center, in the frame in which the rotational-rotational diffusion matrix is diagonal, and compare with 
our theoretical expressions for the time-dependent cross-correlation matrix $\mathbf{C}(t)$ for the mobility center $\mathbf{R}_C(t)$. 

In contrast to numerical simulations, the analytical expressions provided in this work allow to determine the cross-correlations exactly. The accuracy of theoretical expressions is especially important for times comparable to the characteristic time scales of the rotational diffusion, when the cross-correlations change significantly with time. 

\acknowledgments
M.L.E.-J. and E.W. were supported in part by the Polish National Science Centre under Grant No. 2012/05/B/ST8/03010. M.L.E.-J. benefited from the scientific activities of the
COST Action~MP1305. \\

\appendix
\section{Change of a second rank tensor due to the rotational diffusion}\label{srt}
In Eqs.~\eqref{Vt}-\eqref{f1}, we derived the time-dependence of a vector, which rotates with the particle. Now we will do the same for a second rank tensor.

Let $\boldsymbol{H}(t)$ be any second rank tensor which rotates with the
particle, with $\boldsymbol{H}\equiv \boldsymbol{H}(0)$. From Eq. \eqref{BBB} it follows that time evolution of this tensor due to rotational diffusion only is given as 
\begin{equation}
\langle \boldsymbol{H}(t)\rangle_0=\exp (\mathcal{L}^{rr}t)\boldsymbol{H}, \label{Ht}
\end{equation}
The explicit expression for $\boldsymbol{H}(t)$ follows from Eqs.  \eqref{rotB} (see
Ref.
\cite{Rallison}). As in the whole paper, we adopt the frame of reference in which the rotational-rotational diffusion tensor is diagonal, as in Eq. \eqref{dia}. For the diagonal components we obtain,
\begin{eqnarray}
\hspace{-1.4cm} &&\langle \boldsymbol{H}(t)\rangle_{0,ii} =\frac{1}{3}\text{tr}\boldsymbol{H~}\label{Htd} 
+\exp (\mathcal{-}f^{(+)}t)\boldsymbol{H}^{(+)}_{ii} \notag \\
\hspace{-1.4cm} &&+\exp (\mathcal{-}f^{(-)}t)\boldsymbol{H}^{(-)}_{ii},  \label{deH}
\end{eqnarray}
where the first term is a scalar, and as such, it does not change in time, and the characteristic exponents $f^{(+)}$ and $f^{(-)}$ are given in Eqs.~\eqref{f+}-\eqref{f-}.

For the off-diagonal components $i\ne j$,
\begin{eqnarray}
\hspace{-2.4cm}\!\!\! &&\langle \boldsymbol{H}(t)\rangle_{0,ij} =\exp (\mathcal{-}f_{k }^{(1)}t)\boldsymbol{H}^{(a)}_{ij} + \exp (\mathcal{-}f_{k }^{(2)}t)\boldsymbol{H}^{(s)}_{ij} \notag \\ 
\label{Hsa}
\end{eqnarray}
where $k$ is the remaining third index, $k\ne i,j$, and the characteristic exponents $f_i^{(1)}$ and $f_j^{(2)}$ are given in Eqs.~\eqref{f1}-\eqref{f2}.

In Eq.~\eqref{Hsa}, we use the decomposition of 
a second rank tensor $\boldsymbol{H}$ into symmetric and antisymmetric parts, defined as
\bee
&&\left[ \boldsymbol{H}^{(s)}\right] _{ij}=\frac{1}{2}\left(
H_{ ij }+H_{ji }\right), \label{sym}
\\
&&\left[ \boldsymbol{H}^{(a)}\right] _{ij}=\frac{1}{2}\left(
H_{ij }-H_{ji }\right).\label{anti}
\eee

In Eq.~\eqref{deH}, the symmetric diagonal part is further split into 
the isotropic part,
$\frac{1}{3}\text{tr}\boldsymbol{H~}%
\delta _{ij}$, 
and the diagonal non-isotropic parts,
\bee
\left[ \boldsymbol{H}^{(\pm )}\right] _{i i }\!\!\!\!&=&\!\!\left( \!\frac{1}{2}%
\mp \frac{3}{4}\frac{D_{i }\!-\!D}{\Delta }\!\right) \left(\!
H_{i i }\!-\!\frac{1}{3}\text{tr}\boldsymbol{H}\!\right) \nonumber \\
&\pm& \frac{%
D_{j }-D_{k }}{4\Delta }\left( H_{j j }-H_{k k
}\right).
\eee
Above, $j$ and $k\ne j $ are the remaining indices different than $i$.\\\\

\section{Rotational-translational correlations}\label{A}

Using Eq. \eqref{defDeltau}, one can write
\begin{widetext}
\begin{eqnarray}
\left\langle  \mathbf{u}^{(p)}(t) \Delta\mathbf{R}_C(t)%
\right\rangle_0 =
\!\int \!\!d\mathbf{\Omega} \;
\mathbf{u}^{(p)}(\mathbf{\Omega})
\!\int\!\! d%
\mathbf{R}_C\ (\mathbf{R_C\!-\!R}_{C,0})P(\mathbf{R}_C,\mathbf{\Omega }\left\vert 
\mathbf{R}_{C,0},\mathbf{\Omega }_{0};t\right). \label{dd} 
\end{eqnarray}
\end{widetext}

From the Smoluchowski equation, with the use of Eq.~\eqref{Azero}, it follows that in the mobility center, where $\bm{D}^{rt}$ is symmetric, one has, 
\begin{eqnarray}
&&\hspace{-0.5cm}\frac{\partial }{\partial t}\int d\mathbf{R}_C\ (\mathbf{R}_C-\mathbf{R}_{C,0})P(\mathbf{R}_C, 
\mathbf{\Omega }\left\vert \mathbf{R}_{C,0},\mathbf{\Omega }_{0};t\right)\nonumber \\ &&\hspace{-0.5cm}=
{\cal L}^{rr}\!\!\int d\mathbf{R}_C\ (\mathbf{R}_C-\mathbf{R}%
_{C,0})P(\mathbf{R}_C,\mathbf{\Omega }\left\vert \mathbf{R}_{C,0},\mathbf{\Omega }%
_{0};t\right)\!\!\nonumber \\ &&\hspace{-0.5cm}
- 2\frac{\partial }{\partial \bm{\alpha }}\cdot  \bm{D}^{rt}\,
P(\mathbf{
\Omega }\left\vert \mathbf{\Omega }_{0};t\right), 
\label{sko}
\end{eqnarray}
with $P(\widehat{\mathbf {\Omega }}\vert \widehat{\mathbf{\Omega }}_{0};t)$ defined in Eq.~\eqref{redP}. In Eq.~\eqref{sko}, the terms which contain $\frac{\partial }{\partial \bm{\alpha }}\cdot  \bm{D}^{rt}$ and $\bm{D}^{tr} \cdot \frac{\partial }{\partial \bm{\alpha }}$ give identical contributions.

Solving Eq.~\eqref{sko}, we obtain 

\bee
&&\hspace{-0.5cm}\int d\mathbf{R}_C\ (\mathbf{R_C\!-\!R}_{C,0})P(\mathbf{R_C},\mathbf{\Omega }\left\vert 
\mathbf{R}_{C,0},\mathbf{\Omega }_{0};t\right) \nonumber \\
&&\hspace{-0.5cm}= -
2\int\limits_{0}^{t}d\tau
\exp [\mathcal{L}^{rr}(t\!-\!\tau )]
 \frac{\partial }{%
\partial \bm{\alpha }}\cdot \bm{D}^{rt}
\, P\!\left(\mathbf{\widehat{\Omega }}\right.\left\vert \widehat{\mathbf{\Omega }}_{0};\tau\right)\!.
\nonumber \\ \label{aB}\eee
We now insert the expression \eqref{aB} into Eq.~\eqref{dd}, take into account that the Smoluchowski operator is self-adjoint, and perform the integration by parts with respect to $\bm{\alpha}$. We benefit from choosing the frame of reference, in which $\bm{D}^{rr}$ is diagonal, with
\bee
{\cal L}^{rr} \mathbf{u}^{(p)} = - f^{(1)}_p \mathbf{u}^{(p)},
\eee 
and we use the relations \eqref{vecten}, \eqref{rotB} to write for a vector $\bm{V}$ and a matrix $\bm{A}$, 
\bee
\frac{\partial {V}_m}{\partial {\alpha}_k}  {A}_{kn}= - \left(\mathbf{V} \times \bm{S} \right)_{mn}.
\eee
Finally, using Eq. \eqref{BBB}, we obtain Eq.~\eqref{78}.

\section{Shift of the reference center}\label{shift}
The correlations $
\left\langle \Delta \boldsymbol{u}(t)\Delta   \boldsymbol{u}(t) \right\rangle_0 $ do not depend on the choice of the reference center. 
Indeed, they depend only on the rotational-rotational components of the diffusion tensor (or, equivalently, the mobility matrix). But these components do not depend on the choice of a reference center, as it follows from Eqs.~\eqref{trans1}-\eqref{trans3}.

In this appendix, we derive expressions which allow to transform the correlations of the rotational and translational Brownian displacements, $\left\langle \Delta \boldsymbol{u}(t)\Delta   \mathbf{R}(t) \right\rangle_0 $, from one reference center $\mathbf{R}%
_{1}$ to another, $\mathbf{R}%
_{2}$. We denote,
\begin{equation}
\mathbf{R}_{2}=\mathbf{R}_{1}+\mathbf{R}_{21}.
\end{equation}
The difference $\mathbf{R}_{21}$ rotates with the particle, and therefore it can be 
expressed using the set of the orientation vectors $\boldsymbol{u}^{(p)}$, with $p=1,2,3$,
\begin{equation}
\mathbf{R}_{21}=\sum_{p=1}^{3}a_{p}\ \boldsymbol{u}^{(p)}.
\end{equation}%
Therefore,
\begin{equation}
\Delta \mathbf{R}_{2}(t)=\Delta \mathbf{R}%
_{1}(t)+\sum_{p=1}^{3}a_{p} \left(\boldsymbol{u}^{(p)}(t)-\boldsymbol{u}^{(p)}(0)\right)      %
\end{equation}%
Since $\langle \Delta \boldsymbol{u}(t)\rangle_0=0$, 
the transformation of the rotational-translational correlations between the reference centers $\mathbf{R}%
_{1}$ and $\mathbf{R}%
_{2}$ has the form,
\bee &&
\left\langle \Delta \boldsymbol{u}(t)\Delta \mathbf{R}%
_{2}(t)\right\rangle_0 =\left\langle \Delta \boldsymbol{u}%
(t)\Delta \mathbf{R}_{1}(t)\right\rangle_0
+\nonumber \\
&&\sum_{p=1}^{3}a_{p}\left\langle \Delta \boldsymbol{u}(t)
\boldsymbol{u}^{(p)}(t)\right\rangle_0.\label{couR}
\eee
To determine the difference, we
take into account that $\Delta \boldsymbol{u}(t)$ is proportional to 
$\bm{\mathcal R}^{(a)}$\!, see Eq.~\eqref{DeltauRa}, while according to Eq. \eqref{natR}, $\boldsymbol{u}^{(p)}(t)$ contains both $\bm{\mathcal R}^{(a)}$ and $\bm{\mathcal R}^{(s)}$,
\bee
\boldsymbol{u}^{(p)}(t)= \bm{\mathcal R}^{(a)}(t) \cdot \boldsymbol{u}^{(p)}(0) + \bm{\mathcal R}^{(s)}(t) \cdot \boldsymbol{u}^{(p)}(0). \;\;\;\;
\eee 
However, as shown in Eq.~\eqref{asde}, the symmetric and antisymmetric parts of the rotation matrix are not correlated. Therefore, owing to Eq. \eqref{DeltauRa}, the only contribution to $ \boldsymbol{u}^{(p)}(t)$ in Eq.~\eqref{couR} comes from 
\bee
\bm{\mathcal R}^{(a)}(t)\cdot \boldsymbol{u}^{(p)}(0) = \Delta \boldsymbol{u}%
(t) \times \boldsymbol{u}^{(p)}(0).\\\nonumber
\eee
Therefore,
\bee
\!\left\langle \Delta \boldsymbol{u}(t) \boldsymbol{u}%
^{(p)}(t)\right\rangle_0\!\! =\!\left\langle \Delta \boldsymbol{u}%
(t)\Delta \boldsymbol{u}(t)\right\rangle_0 \times \boldsymbol{u}^{(p)}(0),\;\;\;\;\;
\eee%
and we finally obtain the transformation relation,%
\bee &&
\left\langle \Delta \boldsymbol{u}(t)\Delta \mathbf{R}%
_{2}(t)\right\rangle_0 =\left\langle \Delta \boldsymbol{u}%
(t)\Delta \mathbf{R}_{1}(t)\right\rangle_0 +\nonumber \\
&&\left\langle
\Delta \boldsymbol{u}(t)\Delta \boldsymbol{u}%
(t)\right\rangle_0 \times \mathbf{R}_{21}(0).
\eee

This relation can be used to transform the expressions \eqref{rtaa}-\eqref{rtab}, valid for the mobility center $\mathbf{R}_{1}\!=\!\mathbf{R}_{C}$, to account for the rotational-translational correlations 
determined for an arbitrary center $\mathbf{R}_{2}$. 

The analogical transformation relations for the translational-translational correlations $\left\langle \Delta \mathbf{R}(t) \Delta \mathbf{R}%
(t)\right\rangle_0$ contain also correlations involving 
$\bm{\mathcal R}^{(s)}(t)$, and therefore they cannot be expressed only in terms of the correlations ${\bf C}(t)$, and as such, they are not practical if only ${\bf C}(t)$ is measured. However, these additional correlations, $\left\langle \Delta \mathbf{R}(t) \boldsymbol{u}^{(p)}(t)\right\rangle_0$ and $\left\langle  \boldsymbol{u}^{(p)}(t) \boldsymbol{u}^{(q)}(t)\right\rangle_0$, can be also determined in experiments. In this case, the corresponding expressions would be useful and can be easily derived based on the framework constructed in this work. 



\begin{thebibliography}{99}
\bibitem{2D} A. Chakrabarty, A. Konya, F. Wang, J. V. Selinger, K. Sun, and Qi-Huo Wei, Langmuir {\bf 30}, 13844 
(2014).

\bibitem{holografic} A. Wang, T. G. Dimiduk, J. Fung, S. Razavi, I. Kretzschmar, K. Chaudhary, and V. N. Manoharan, J. Quant. Spectrosc. Radiat. Transfer {\bf 146}, 499 
(2014). 

\bibitem{leptospira} L. Koens and E. Lauga, Phys. Biol. {\bf 11}, 066008 (2014).

 
\bibitem{Adamczyk} Z. Adamczyk, B. Cichocki, M. L. Ekiel-Je\.zewska, A. S\l owicka,
E. Wajnryb, and M. Wasilewska, 
J. Colloid Interface Sci. {\bf 385}  244 (2012).

\bibitem{Winkler} R. G. Winkler, J. Chem. Phys. {\bf 133}, 164905 (2010).

\bibitem{Goldstein} V. Kantsler, R. E. Goldstein, Phys. Rev. Lett. {\bf 108}, 038103
(2012).

\bibitem{Kraft}
D. J. Kraft, R. Wittkowski, B. ten Hagen, K. V. Edmond, D. J. Pine and H. L\"owen, Phys. Rev. E {\bf 88}, 050301 (2013). 

\bibitem{Kampen} N. van Kampen, {\it Stochastic Processes in Physics and Chemistry, 3rd Edition}, North-Holland, 2007.

\bibitem{Rallison}
J. M. Rallison, J. Fluid Mech. {\bf 84} 237 (1978).

\bibitem{CEJW_2012intrinsic}
B. Cichocki, M. L. Ekiel-Je\.zewska and E. Wajnryb, 
J. Phys: Conference Series, {\bf 392}, 012004 (2012).

\bibitem{CEJW_2012trans}
B. Cichocki, M. L. Ekiel-Je\.zewska and E. Wajnryb, {J. Chem. Phys.} {\bf 136}, 071102 (2012).

\bibitem{SE1}
M. Doi and S. F. Edwards, {\it The Theory of Polymer Solutions}, Clarendon,
Oxford (1988).

\bibitem{SE2}
R. B. Jones and P. N. Pusey, Annu. Rev. Phys. Chem. {\bf 42}, 137 (1991).

\bibitem{SE3}
J. K. G. Dhont, {\it An Introduction to Dynamics of Colloids}, Elsevier,
Amsterdam (1996).

\bibitem{Favro}
L. D. Favro, Phys. Rev. {\bf 119} 53 (1960).

\bibitem{BernePecora}
B. Berne and R. Pecora, {\it Dynamic Light Scattering: With Applications to Chemistry, Biology and Physics}, 
New York, Wiley (1976).

\bibitem{BruneKim:1993}
D.~Brune and S.~Kim,
\newblock {Proc. Natl. Acad. Sci. USA} {\bf 90}, 3835 (1993).

\bibitem{KimKarrila}
S. Kim and S. J. Karrila,
\newblock {\em Microhydrodynamics. Principles and Selected Applications}.
\newblock Dover Publications, Mineola, 2005.

\bibitem{Euler} H. Goldstein, C. P. Poole, J. L. Safko, {\it Classical Mechanics}, Third Edition, Addison-Wesley, 2001. 

\end{thebibliography}
\end{document}